\documentclass[12pt]{article}

\usepackage[T1]{fontenc}
\usepackage{lmodern}

\usepackage{setspace}
\setdisplayskipstretch{1}

\usepackage[margin=1.25in]{geometry}

\usepackage[dvipsnames]{xcolor}
\usepackage{pdfpages}
\usepackage{float}
\usepackage{multirow}
\usepackage{caption}
\usepackage{subcaption}
\usepackage{epigraph}
\usepackage{titlesec}

\titleformat{\section}
  {\normalfont\large\bfseries}{\thesection}{1em}{}
\titleformat{\subsection}
  {\normalfont\normalsize\bfseries}{\thesubsection}{1em}{}

\usepackage{mathtools}
\usepackage{amssymb}
\usepackage{amsthm}
\usepackage{amsfonts}
\usepackage{aliascnt}
\usepackage{accents}
\usepackage{dutchcal}
\usepackage{bbm}

\usepackage{enumitem}
\usepackage{sgame}
\usepackage{tikz}
\usepackage{tikz-cd}
\usetikzlibrary{calc,shapes,arrows}
\usepackage{tcolorbox}
\usepackage{listings}
\usepackage[normalem]{ulem}

\usepackage[round]{natbib}
\DeclareMathOperator{\supp}{supp}

\DeclareMathOperator*{\argmax}{arg\,max}

\newcommand{\BN}{\mathsf{BN}}

\newtheorem{theorem}{Theorem}[section]

\newaliascnt{proposition}{theorem}
\newtheorem{proposition}[proposition]{Proposition}
\aliascntresetthe{proposition}

\newaliascnt{lemma}{theorem}
\newtheorem{lemma}[lemma]{Lemma}
\aliascntresetthe{lemma}

\newaliascnt{corollary}{theorem}
\newtheorem{corollary}[corollary]{Corollary}
\aliascntresetthe{corollary}

\newaliascnt{claim}{theorem}
\newtheorem{claim}[claim]{Claim}
\aliascntresetthe{claim}

\theoremstyle{definition}

\newaliascnt{definition}{theorem}

\aliascntresetthe{definition}

\newaliascnt{example}{theorem}

\aliascntresetthe{example}

\newaliascnt{assumption}{theorem}

\aliascntresetthe{assumption}

\newtheorem{axiom}{Axiom}

\newaliascnt{condition}{theorem}

\aliascntresetthe{condition}

\newaliascnt{question}{theorem}

\aliascntresetthe{question}

\newaliascnt{remark}{theorem}

\aliascntresetthe{remark}

\newaliascnt{remarks}{theorem}

\aliascntresetthe{remarks}

\newaliascnt{aside}{theorem}

\aliascntresetthe{aside}

\newaliascnt{note}{theorem}

\aliascntresetthe{note}

\usepackage{thmtools}
\usepackage{thm-restate}

\usepackage{hyperref}

\hypersetup{
    colorlinks=true,
    linkcolor=OrangeRed,
    filecolor=Thistle,
    urlcolor=Thistle,
    citecolor=Thistle,
}

\usepackage[nameinlink]{cleveref}

\crefname{theorem}{theorem}{theorems}
\Crefname{theorem}{Theorem}{Theorems}
\crefname{proposition}{proposition}{propositions}
\Crefname{proposition}{Proposition}{Propositions}
\crefname{lemma}{lemma}{lemmas}
\Crefname{lemma}{Lemma}{Lemmas}
\crefname{corollary}{corollary}{corollaries}
\Crefname{corollary}{Corollary}{Corollaries}
\crefname{claim}{claim}{claims}
\Crefname{claim}{Claim}{Claims}
\crefname{definition}{definition}{definitions}
\Crefname{definition}{Definition}{Definitions}
\crefname{example}{example}{examples}
\Crefname{example}{Example}{Examples}
\crefname{assumption}{assumption}{assumptions}
\Crefname{assumption}{Assumption}{Assumptions}
\crefname{axiom}{axiom}{axioms}
\Crefname{axiom}{Axiom}{Axioms}

\makeatletter
\let\cref@old@isrefconsecutive\cref@isrefconsecutive
\def\cref@isrefconsecutive#1#2{%
  \begingroup
    \def\cref@assumptiontype{assumption}%
    \cref@gettype{#1}{\cref@typea}%
    \ifx\cref@typea\cref@assumptiontype
      \endgroup
      \@cref@refconsecutivefalse
    \else
      \endgroup
      \cref@old@isrefconsecutive{#1}{#2}%
    \fi
}
\makeatother

\crefname{condition}{condition}{conditions}
\Crefname{condition}{Condition}{Conditions}
\crefname{question}{question}{questions}
\Crefname{question}{Question}{Questions}
\crefname{remark}{remark}{remarks}
\Crefname{remark}{Remark}{Remarks}
\crefname{remarks}{remarks}{remarks}
\Crefname{remarks}{Remarks}{Remarks}
\crefname{aside}{aside}{asides}
\Crefname{aside}{Aside}{Asides}
\crefname{note}{note}{notes}
\Crefname{note}{Note}{Notes}
\crefname{appendix}{appendix}{appendices}
\Crefname{appendix}{Appendix}{Appendices}

\newcommand{\secref}[1]{\hyperref[#1]{\S\ref*{#1}}}

\definecolor{backcolour}{rgb}{0.63,0.79,0.95}
\lstdefinestyle{mystyle}{
  backgroundcolor=\color{backcolour},
  basicstyle=\ttfamily\footnotesize,
  breakatwhitespace=false,
  breaklines=true,
  captionpos=b,
  keepspaces=true,
  numbers=left,
  numbersep=5pt,
  showspaces=false,
  showstringspaces=false,
  showtabs=false,
  tabsize=2
}
\begin{document}
\title{How to Pick a Model if You Must}
\author{Mark Whitmeyer\thanks{Thanks to Joseph Whitmeyer, ChatGPT, Deep Review, and \href{https://www.refine.ink/}{refine.ink} for help. For KS.}}
\date{\today}
\maketitle
\begin{abstract}
I axiomatize Berk-Nash equilibrium in decision problems with possibly misspecified models. I specify how a rule--which selects a set of action-belief pairs in a problem--ought to compare a correct forecast with an incorrect one, react to the removal of models, and accommodate additional observations that do not distinguish among the models. Together with axioms concerning attitudes toward randomization and payoffs, and a pair of regularity conditions, these axioms characterize Berk-Nash equilibrium.
\end{abstract}

\section{Being firm}

A firm is choosing what price to set. It has various models of demand at its disposal, each of which specifies a probability distribution over sales at every price. The firm uses these forecasts to compute prospective profits and compares the predictions with the sales it observes to decide which models to retain. How should it do so?

The firm's behavior is governed by an \textit{assessment}, which consists of a pricing policy--a probability distribution over prices--and a probability distribution over models. A \textit{rule} selects a set of assessments for each specification of payoffs, sales distributions, and models. How should such rules behave?

First suppose that the firm can charge only one price and that its utility is zero regardless of how much it sells. The rule then selects beliefs about demand alone. If one of two models predicts the distribution of sales correctly and the other does not, every belief selected by the rule should assign probability one to the correct model (\Cref{ax:EF}).

How should a firm handle the removal of models from consideration? Suppose at least one of two models receives positive probability from a belief that the rule selects when all models are available. Removing all other models should leave unchanged which of the two can receive positive probability (\Cref{ax:SM}). After all, the two remaining models still make the same predictions about the same sales observations.

Suppose next that each sales observation comes with an administrative report that has no effect on profits. Conditional on the price and sales, all models assign the same probability to each report that can occur under the firm's policy (they may disagree about reports that cannot occur--but who cares about them). Observing this report should not affect whether the rule selects the firm's policy and belief (\Cref{ax:IF}). The common prediction may be wrong, and such an error can count against all the models, but it should not single out one of them. Each model's original sales forecast remains fixed at every price.

The other properties concern the firm's choices. It should not matter whether the firm randomizes directly over prices or chooses a lottery that produces the same distribution over prices, provided the firm observes the price drawn (\Cref{ax:RR}). For a price chosen with certainty and a fixed belief, selection of that price and belief should be unaltered when, at any price, the payoffs to quantities sold regarded as equally likely under that belief are exchanged (\Cref{ax:PS}). If the rule selects the same price and belief under each of two utility functions, it ought also to select them when we average the two utilities assigned to each amount of sales at each price (\Cref{ax:PC}).

A firm should never choose for sure a price whose best payoff is below another price's worst payoff (\Cref{ax:SP}). Finally, if the rule selects a price and belief, it should still select that price and belief when all utilities are halved (\Cref{ax:HS}).

My main result, \Cref{thm:joint}, is that among rules that always select at least one assessment and have a closed graph, these axioms pin down the rule that selects exactly the Berk-Nash assessments \citep[Definition~1]{esponda2016}.

\section{Fitting models}\label{sec:model-selection}

There is a finite nonempty set of outcomes $Z$ and a finite nonempty set of models $\Theta$. The true distribution of outcomes is $P\in\Delta(Z)$. Each model $\theta\in\Theta$ specifies a full-support distribution $Q_\theta\in\Delta^\circ(Z)$ over outcomes. Write $Q=(Q_\theta)_{\theta\in\Theta}$. Given a finite nonempty set \(S\), a \emph{model-selection rule} assigns a nonempty set $C(\tilde P,\tilde Q)\subseteq\Theta$ to each distribution $\tilde P\in\Delta(S)$ and each family $\tilde Q=(\tilde Q_\theta)_{\theta\in\Theta}$ (where \(\tilde Q_\theta \in \Delta^\circ(S)\)).

For $N\subseteq\Theta$, write $Q|_N \coloneqq (Q_\theta)_{\theta\in N}$. For a finite set of \textit{observations} \(W\) (with \(|W| \geq |Z|\)) let $\rho\colon W\to Z$ be onto. For each $z\in Z$, let $T(\cdot\mid z)\in\Delta(\rho^{-1}(z))$ be the true conditional distribution of observations and let $R_\theta(\cdot\mid z)\in\Delta^\circ(\rho^{-1}(z))$ be the conditional distribution of observations specified by model $\theta$. The induced distributions over observations are:
\begin{equation}\label{eq:feedback} \widehat P(w)=P(\rho(w))T(w\mid\rho(w)) \qquad \text{\&} \qquad \widehat Q_\theta(w)=Q_\theta(\rho(w))R_\theta(w\mid\rho(w)). \end{equation}

I now state three desirable properties for a model-selection rule--I introduce decision-making into the fray in \secref{sec:joint}. The first axiom governs how selection responds when models are removed, the second fixes selection when one of two models is correct, and the third requires invariance to additional feedback that every model predicts in the same way conditional on the original outcome.
\begin{enumerate}[label=(\roman*),nosep,noitemsep]
\item \textbf{Model-menu consistency.} For $N\subseteq\Theta$ with \(|N| = 2\),
\begin{equation}\label{eq:model-menu} C(P,Q)\cap N\ne\emptyset \quad\Longrightarrow\quad C(P,Q|_N)=C(P,Q)\cap N. \end{equation}
\item \textbf{Binary exact fit.} If $\Theta=\{\theta^*,\phi\}$, $Q_{\theta^*}=P$, and $Q_\phi\ne P$, then
\begin{equation}\label{eq:exact-fit} C(P,Q)=\{\theta^*\}. \end{equation}
\item \textbf{Irrelevant feedback.} If $R_\theta(w\mid\rho(w))$ is the same for every model whenever $\widehat P(w)>0$, then
\begin{equation}\label{eq:feedback-invariance} C(\widehat P,\widehat Q)=C(P,Q). \end{equation}
\end{enumerate}

\eqref{eq:model-menu} compares the rule's selection when all models in $\Theta$ are available with its selection when only the two models in $N$ are available, saying that if the rule selects at least one member of $N$ when all models are available, removing the models outside $N$ does not change which members of $N$ it selects. \eqref{eq:exact-fit} then means that given a choice between two models, one of which is correct and the other incorrect, the rule selects only the correct model. Finally, \eqref{eq:feedback-invariance} says that using additional feedback does not change which models the rule selects if, conditional on the original outcome, all models assign the same probability to every observation that occurs with positive true probability.

A rule $C$ has a \emph{closed graph} if, for fixed $S$ and $\Theta$, $(\tilde P^k,\tilde Q^k)\to(\tilde P,\tilde Q)$ and $\theta\in C(\tilde P^k,\tilde Q^k)$ for every $k$ imply $\theta\in C(\tilde P,\tilde Q)$, provided every distribution in $\tilde Q$ has full support. Among such rules, the three axioms above characterize the rule that selects exactly the models maximizing expected log likelihood.

\begin{proposition}\label{prop:inference}
A model-selection rule $C$ with a closed graph satisfies \eqref{eq:model-menu}, \eqref{eq:exact-fit}, and \eqref{eq:feedback-invariance} if and only if
\begin{equation}\label{eq:inference-target} C(P,Q)=\argmax_{\theta\in\Theta}\sum_{z\in Z}P(z)\log Q_\theta(z). \end{equation}
\end{proposition}

\section{Wherefore Berk-Nash}\label{sec:joint}

There is a finite nonempty set of actions $A$, a finite nonempty set of models $\Theta$, and a finite nonempty set of outcomes $Z_a$ for each action $a$. The true distribution of outcomes after action $a$ is $P_a\in\Delta(Z_a)$. Each model $\theta$ specifies a full-support distribution $Q_{\theta a}\in\Delta^\circ(Z_a)$. The payoff from outcome $z$ after action $a$ is $u_a(z)\in\mathbb R$. Write $u=(u_a)_{a\in A}$, $P=(P_a)_{a\in A}$, and $Q=(Q_{\theta a})_{\theta\in\Theta,a\in A}$. A \textit{problem} is $E=(u,P,Q)$.

An \emph{assessment} is a pair $(\alpha,\nu)\in\Delta(A)\times\Delta(\Theta)$, where $\alpha$ is a mixed action and $\nu$ is a belief over models. Under belief $\nu$, the predicted distribution at action $a$ is $\overline Q_a^\nu\coloneqq\sum_\theta\nu_\theta Q_{\theta a}$. Write $\delta_a$ for the distribution that assigns probability one to action $a$. An assessment is \emph{pure} if it has the form $(\delta_a,\nu)$ (note that the belief $\nu$ can still put positive probability on several models).

A \textit{rule} $F$ assigns a set $F(E)\subseteq\Delta(A)\times\Delta(\Theta)$ to each problem, for all finite choices of actions, models, and outcomes. It is \emph{total} if $F(E)\ne\emptyset$ for every $E$. It has a \emph{closed graph} if, on fixed sets $A,\Theta,(Z_a)_{a\in A}$, $E^k\to E$, $(\alpha^k,\nu^k)\to(\alpha,\nu)$, and $(\alpha^k,\nu^k)\in F(E^k)$ for every $k$ imply $(\alpha,\nu)\in F(E)$, provided every limiting model distribution has full support.\footnote{Totality ensures that the rule selects an assessment even when every model is wrong, whereas a closed graph ensures that a limit of selected assessments remains selected as the problem converges.}

For each action and model, define expected payoff and expected log likelihood by
\begin{equation}\label{eq:payoffs-and-likelihoods} m_{a\theta}\coloneqq\sum_zQ_{\theta a}(z)u_a(z) \qquad \text{\&} \qquad \ell_{a\theta}\coloneqq\sum_zP_a(z)\log Q_{\theta a}(z). \end{equation}
The Berk-Nash correspondence is
\begin{equation}\label{eq:BN} \BN(E)=\left\{(\alpha,\nu)\colon \supp\alpha\subseteq\argmax_a\sum_\theta\nu_\theta m_{a\theta},\ \supp\nu\subseteq\argmax_\theta\sum_a\alpha_a\ell_{a\theta}\right\}. \end{equation}
The first condition says that every action used with positive probability maximizes expected payoff under $\nu$. The second says that every model assigned positive probability maximizes expected log likelihood under $\alpha$. I refer to these as the \emph{choice condition} and the \emph{fitting condition}, respectively. This is \citet[Definition~1]{esponda2016} for a single decision maker with finite actions, models, and outcomes.

Consider a new problem with a finite set of choices $B$. Each choice $b\in B$ implements a specified lottery $\lambda_b\in\Delta(A)$ over the original actions. Require that, for every $a\in A$, there is some $b\in B$ with $\lambda_b=\delta_a$, so every original action can still be chosen with certainty. Choosing $b$ draws an original action $a$ with probability $\lambda_b(a)$. Conditional on this draw, the outcome $z\in Z_a$ has true distribution $P_a$, and model $\theta$ predicts $Q_{\theta a}$. The decision maker observes both $a$ and $z$. Thus, the set of outcomes is $\widehat Z_b\coloneqq\{(a,z)\colon\lambda_b(a)>0, z\in Z_a\}$, the true distribution is $\widehat P_b(a,z)\coloneqq\lambda_b(a)P_a(z)$, model $\theta$ predicts $\widehat Q_{\theta b}(a,z)\coloneqq\lambda_b(a)Q_{\theta a}(z)$, and the payoff is $\widehat u_b(a,z)\coloneqq u_a(z)$. Write $\widehat E=(\widehat u,\widehat P,\widehat Q)$ for this problem. A distribution $\widehat\alpha\in\Delta(B)$ over the new choices induces the distribution $\sum_{b\in B}\widehat\alpha_b\lambda_b$ over the original actions.

\begin{axiom}\label{ax:RR}
For every $(\widehat\alpha,\nu)\in\Delta(B)\times\Delta(\Theta)$, \((\widehat\alpha,\nu)\in F(\widehat E) \quad\Longleftrightarrow\quad \left(\sum_{b\in B}\widehat\alpha_b\lambda_b,\nu\right)\in F(E)\).
\end{axiom}
\Cref{ax:RR} requires the rule to select a lottery and a belief if and only if it selects that belief and the distribution over actions that the lottery produces. Suppose, for example, that choice $b$ draws actions $a_1$ and $a_2$ with equal probability. Choosing $b$ with certainty and choosing each of $a_1$ and $a_2$ with probability one half produce the same distribution over actions. The rule must select both with belief $\nu$, or select neither with that belief. The requirement also applies when the decision maker mixes over several lotteries: the overall probability of taking action $a$ is then $\sum_{b\in B}\widehat\alpha_b\lambda_b(a)$, and the comparison is with directly choosing the original actions according to these probabilities.

The next axiom concerns additional information observed along with the original outcome. For each action $a$, let $W_a$ be a finite set of observations and let $\rho_a\colon W_a\to Z_a$ be onto. An observation $w\in W_a$ records the original outcome $\rho_a(w)$ and may contain further information. In that light, $\rho_a^{-1}(z)$ is the set of observations that record outcome $z$. For each $z\in Z_a$, let $T_a(\cdot\mid z)\in\Delta(\rho_a^{-1}(z))$ be the true conditional distribution of observations given $z$, and let $R_{\theta a}(\cdot\mid z)\in\Delta^\circ(\rho_a^{-1}(z))$ be the conditional distribution specified by model $\theta$. Applying the construction in \eqref{eq:feedback} separately to each action, define $\widehat P_a(w)\coloneqq P_a(\rho_a(w))T_a(w\mid\rho_a(w))$ and $\widehat Q_{\theta a}(w)\coloneqq Q_{\theta a}(\rho_a(w))R_{\theta a}(w\mid\rho_a(w))$. Payoffs continue to depend only on the original outcome: $\widehat u_a(w)\coloneqq u_a(\rho_a(w))$. Write $\widehat E=(\widehat u,\widehat P,\widehat Q)$ for this problem.

\begin{axiom}\label{ax:IF}
For every assessment $(\alpha,\nu)$, if $R_{\theta a}(w\mid\rho_a(w))$ is the same for every model whenever $\alpha_a\widehat P_a(w)>0$, then \((\alpha,\nu)\in F(\widehat E) \quad\Longleftrightarrow\quad (\alpha,\nu)\in F(E)\).
\end{axiom}

\begin{axiom}\label{ax:PS}
Fix a belief $\nu$. For each action $b$, let $\sigma_b$ be a permutation of $Z_b$ such that $\overline Q_b^\nu(\sigma_b(z))=\overline Q_b^\nu(z)$ for every $z$, and define $v_b(z)\coloneqq u_b(\sigma_b(z))$. Then, for every $a\in A$,
\[(\delta_a,\nu)\in F(u,P,Q) \quad\Longleftrightarrow\quad (\delta_a,\nu)\in F(v,P,Q).\]
\end{axiom}

\begin{axiom}\label{ax:PC}
For any payoff functions $u$ and $v$, if $(\delta_a,\nu)$ belongs to both $F(u,P,Q)$ and $F(v,P,Q)$, then it belongs to $F((u+v)/2,P,Q)$.
\end{axiom}

\begin{axiom}\label{ax:SP}
If $\min_{z\in Z_b}u_b(z)>\max_{z\in Z_a}u_a(z)$, then $(\delta_a,\nu)\notin F(u,P,Q)$ for every belief $\nu$.
\end{axiom}

\begin{axiom}\label{ax:HS}
If $(\delta_a,\nu)\in F(u,P,Q)$, then $(\delta_a,\nu)\in F(u/2,P,Q)$.
\end{axiom}
Utility is measured in an arbitrary unit. Expressing utility in units twice as large replaces $u$ by $u/2$ and leaves all forecasts unchanged. Under \Cref{ax:HS}, a selected price and belief remain selected after this change of units.

The last two axioms concern problems with a single action $a$ and $u_a(z)=0$ for every outcome. There is no choice of action, and every belief leads to the same payoff. Drop the action subscript from $P_a$ and $Q_{\theta a}$, and define
\begin{equation}\label{eq:SC} S_F(P,Q)\coloneqq\{\nu\colon (\delta_a,\nu)\in F(0,P,Q)\}, \qquad \text{\&} \qquad C_F(P,Q)\coloneqq\bigcup_{\nu\in S_F(P,Q)}\supp\nu. \end{equation}
The set of selected beliefs is $S_F(P,Q)$, and $C_F(P,Q)$ is the set of models that receive positive probability in at least one selected belief. For rules satisfying \Cref{ax:RR,ax:IF}, these sets do not depend on the name of the single action.

\begin{axiom}\label{ax:SM}
For every two-element set $N\subseteq\Theta$, \(C_F(P,Q)\cap N\ne\emptyset \quad\Longrightarrow\quad C_F(P,Q|_N)=C_F(P,Q)\cap N\).
\end{axiom}
\Cref{ax:SM} applies \eqref{eq:model-menu} to the models that receive positive probability in some selected belief. If one of these models remains in $N$, deleting the others leaves unchanged which members of $N$ can receive positive probability.

\begin{axiom}\label{ax:EF}
In a problem with one action and zero payoffs, if $\Theta=\{\theta^*,\phi\}$, $Q_{\theta^*}=P$, and $Q_\phi\ne P$, then every $\nu\in S_F(P,Q)$ assigns probability one to $\theta^*$.
\end{axiom}
As in \eqref{eq:exact-fit}, this axiom says nothing about the choice between incorrect models. I now state my main result.

\begin{theorem}\label{thm:joint}
The Berk-Nash correspondence is the unique total rule with a closed graph satisfying \Cref{ax:RR,ax:IF,ax:PS,ax:PC,ax:SP,ax:HS,ax:SM,ax:EF}.
\end{theorem}

\section{Proving \texorpdfstring{\Cref{thm:joint}}{thm}}\label{sec:joint-proof}

I prove three lemmas. The first two  establish the two requirements in \eqref{eq:BN}, with the first producing the fitting condition and the second the choice condition. The third lemma shows that a total, closed rule satisfying \Cref{ax:RR,ax:IF} cannot select a proper subset of the Berk-Nash assessments.

\begin{lemma}\label{lem:inference-containment}
Every total rule $F$ with a closed graph satisfying \Cref{ax:RR,ax:IF,ax:HS,ax:SM,ax:EF} selects only assessments $(\alpha,\nu)$ in which every model assigned positive probability by $\nu$ maximizes expected log likelihood, with each action's contribution weighted by its probability under $\alpha$.
\end{lemma}

\paragraph{Proof sketch.} Fix $(\alpha,\nu)\in F(u,P,Q)$. Add an action $a$ that first draws an original action $c$ according to $\alpha$ and then produces the outcome generated by $c$. By \Cref{ax:RR}, $(\delta_a,\nu)$ is selected in the expanded problem.

Now halve all payoffs repeatedly. \Cref{ax:HS} preserves selection at every step, and the closed-graph property preserves it in the limit, when all payoffs are zero. The experiment generated by $a$ has observations $(c,z)$ and distributions
\[P^\alpha(c,z)\coloneqq\alpha_cP_c(z) \qquad\text{\&}\qquad Q^\alpha_\theta(c,z)\coloneqq\alpha_cQ_{\theta c}(z).\]
Because no other action is used, two applications of \Cref{ax:IF} let us replace the distributions at each such action by \(P^\alpha,Q^\alpha\): first append $(c,z)$ to the original outcome at that action and then retain $(c,z)$ while treating the original outcome as additional feedback. All actions now have true distribution \(P^\alpha\), model distributions \(Q^\alpha\), and zero payoffs, so \Cref{ax:RR} reduces the problem to one action.

In the resulting one-action problem, model-menu consistency, binary exact fit, and irrelevant feedback imply that every model in $\supp\nu$ maximizes expected log likelihood. For model $\theta$, this expected log likelihood is
\[\sum_{c\in\supp\alpha}\sum_{z\in Z_c} \alpha_cP_c(z)\log\left(\alpha_cQ_{\theta c}(z)\right) = \sum_c\alpha_c\ell_{c\theta} + \sum_{c\in\supp\alpha}\alpha_c\log\alpha_c.\]
The last term does not depend on $\theta$, and so every model in $\supp\nu$ maximizes $\sum_c\alpha_c\ell_{c\theta}$.

\begin{lemma}\label{lem:choice-containment}
Every rule $F$ satisfying \Cref{ax:RR,ax:IF,ax:PS,ax:PC,ax:SP} selects only assessments $(\alpha,\nu)$ in which every action chosen with strictly positive probability maximizes expected utility under $\nu$.
\end{lemma}

\paragraph{Proof sketch.} Suppose instead that $(\alpha,\nu)$ is selected and $\alpha$ assigns positive probability to an action that is not optimal under $\nu$. As in the preceding proof, add a compound action $a$ that draws the original actions according to $\alpha$. By \Cref{ax:RR}, $(\delta_a,\nu)$ is selected in the expanded problem. The expected payoff from $a$ is the $\alpha$-weighted average of their expected payoffs. Because this average puts strictly positive weight on a suboptimal action, some original action $b$ has strictly greater expected payoff than $a$ under $\nu$.

The difficulty is that $b$ does better only on average, whereas \Cref{ax:SP} applies only when every payoff at $b$ exceeds every payoff at $a$. To turn this comparison of expected payoffs into the stronger condition required by \Cref{ax:SP}, we add sufficiently fine observations at the two actions that initially have no effect on payoffs, which split each original outcome into many refined outcomes. We choose the refinement so that, at each action $d\in\{a,b\}$, the relevant permutations of payoffs exchange only refined observations that have the same probability under the $\nu$-weighted distribution over outcomes at $d$. Averaging these permutations makes the payoff at every refined outcome arbitrarily close to the action's expected payoff. \Cref{ax:IF} preserves selection when the observations are added, \Cref{ax:PS} preserves it under each permutation, and repeated applications of \Cref{ax:PC} preserve it under their average.

Since $b$ has strictly greater expected payoff than $a$, a sufficiently fine refinement makes every averaged payoff at $b$ exceed every averaged payoff at $a$. The rule still selects $a$, contradicting \Cref{ax:SP}.

\begin{lemma}\label{lem:no-refinement}
Suppose $F$ is total, has a closed graph, and satisfies \Cref{ax:RR,ax:IF}. If $F(E)\subseteq\BN(E)$ for every problem $E$, then $F=\BN$.
\end{lemma}

\paragraph{Proof sketch.} Fix $(\alpha,\nu)\in\BN(E)$ and add a compound action $a$ that draws the original actions according to $\alpha$. By \Cref{ax:RR}, it is enough to show that $F$ selects $(\delta_a,\nu)$ in the expanded problem.

We approximate this assessment by problems in which it is the unique Berk-Nash assessment. Let $S\coloneqq\supp\nu$. Replace $a$ by one copy $a_j$ for each $j\in S$, and let $\tau$ choose $a_j$ with probability $\nu_j$. Add payoff-irrelevant reports to the unused old actions and slightly perturb both the payoffs and the models' expected log likelihoods for the copies. The perturbations can be chosen so that the Berk-Nash conditions become the two optimality conditions of a zero-sum game. They ensure that any saddle point uses only the copies $a_j$ and the models in $S$. Within this reduced game, under belief $\nu$, the decision maker is indifferent among the copies $a_j$, and, when the copies are chosen according to $\tau$, all models in $S$ fit the resulting observations equally well. These two indifferences uniquely pin down the belief and policy, so $(\tau,\nu)$ is the unique saddle point and hence the unique Berk-Nash assessment of the perturbed problem. Since $F$ is total and selects only Berk-Nash assessments, it must select $(\tau,\nu)$ in every perturbed problem.

As the perturbations disappear, the closed-graph property preserves selection at the limit. \Cref{ax:IF} removes the added reports, while \Cref{ax:RR} first merges the copies into $a$ and then replaces $a$ by the original mixture $\alpha$, yielding $(\alpha,\nu)\in F(E)$.

\begin{proof}[Proof of \Cref{thm:joint}]
Suppose $F$ is total, has a closed graph, and satisfies the eight axioms. By \Cref{lem:inference-containment,lem:choice-containment}, $F(E)\subseteq\BN(E)$ for every problem $E$. Applying \Cref{lem:no-refinement} yields $F=\BN$.

Conversely, the two maximizing correspondences in \eqref{eq:BN} have nonempty, convex values and closed graphs, so \citet[Theorem~2]{kakutani1941} implies totality of $\BN$. Its graph is closed because the inequalities in \eqref{eq:BN} remain valid along converging problems and assessments.

For an action $c$ defined by the lottery $\lambda_c$ in \Cref{ax:RR}, expected payoff is $\sum_a\lambda_c(a)m_{a\theta}$, and expected log likelihood is $\sum_a\lambda_c(a)\ell_{a\theta}+\sum_{a\in\supp\lambda_c}\lambda_c(a)\log\lambda_c(a)$. The last sum is the same for every model. Since every original action remains available, both conditions in \eqref{eq:BN} hold for $(\widehat\alpha,\nu)$ exactly when they hold for $(\sum_c\widehat\alpha_c\lambda_c,\nu)$. Thus, $\BN$ satisfies \Cref{ax:RR}.

In the two problems compared in \Cref{ax:IF}, each model's expected payoff from each action is the same. At each action with $\alpha_a>0$, \eqref{eq:inference-feedback-log-likelihood} implies that the change in expected log likelihood is the same for every model. Consequently, neither the actions maximizing expected payoff under $\nu$ nor the models maximizing $\sum_a\alpha_a\ell_{a\theta}$ change, so $\BN$ satisfies \Cref{ax:IF}. The properties in \Cref{ax:PS,ax:PC,ax:SP,ax:HS} follow from expected-payoff maximization, and changes in payoffs do not affect expected log likelihoods. Finally, in a problem with one action and zero payoffs, $S_{\BN}(P,Q)=\Delta(\argmax_\theta\sum_zP(z)\log Q_\theta(z))$. By \Cref{clm:inference-converse},\footnote{The rule in \eqref{eq:inference-target} selects a nonempty set, has a closed graph, and satisfies \eqref{eq:model-menu}, \eqref{eq:exact-fit}, and \eqref{eq:feedback-invariance}.} $\BN$, therefore, satisfies \Cref{ax:SM,ax:EF}.
\end{proof}

\begin{corollary}\label{cor:derived-coherences}
Every rule satisfying the hypotheses of \Cref{thm:joint} preserves selection when models are renamed and when the probability assigned to a model is divided arbitrarily among identical copies. For every $t>0$, $F(tu,P,Q)=F(u,P,Q)$. For any payoff functions $u,v$,
\[F(u,P,Q)\cap F(v,P,Q)\subseteq F(\lambda u+(1-\lambda)v,P,Q) \quad(0\leq\lambda\leq1).\]
In a problem with one action and zero payoffs, let $N\subseteq\Theta$ be nonempty and define $\nu^N_\theta\coloneqq\nu_\theta/\nu(N)$ on $N$ whenever $\nu(N)>0$. If some selected belief assigns positive probability to $N$, then
\begin{equation}\label{eq:belief-menu} S_F(P,Q|_N) =\{\nu^N\colon \nu\in S_F(P,Q),\ \nu(N)>0\}. \end{equation}
\end{corollary}
\begin{proof}
By \Cref{thm:joint}, the rule equals $\BN$. Renaming models or dividing a model's probability among identical copies preserves both conditions in \eqref{eq:BN}. Multiplying utilities by a positive constant preserves the choice condition. If that condition holds under both $u$ and $v$, it also holds under every convex combination of them. Neither operation changes which models maximize $\sum_a\alpha_a\ell_{a\theta}$. In a problem with one action and zero payoffs, the selected beliefs are exactly those supported on models maximizing expected log likelihood. If $N$ contains a maximizer, its maximizing models are exactly the original maximizing models that belong to $N$. Conditioning the selected beliefs on $N$ thus produces precisely the beliefs supported on those models, which proves \eqref{eq:belief-menu}.
\end{proof}

\section{Independence of the axioms}\label{sec:independence}

\begin{proposition}\label{prop:independence}
Each of \Cref{ax:RR,ax:IF,ax:PS,ax:PC,ax:SP,ax:HS,ax:SM,ax:EF} is independent of the other seven, totality, and the closed-graph property. Totality and the closed-graph property are also each independent of the remaining properties. All counterexamples for the eight axioms are defined on the domain in \Cref{sec:joint}, are total, and have closed graphs.
\end{proposition}

\begin{claim}\label{clm:counterexample-existence}
Suppose a rule selects the pairs in which the action distribution and the model belief each maximize a criterion, holding the other fixed. On fixed finite sets, suppose the criteria are jointly continuous in the problem and both distributions, and that both sets of maximizers are nonempty and convex. Then the rule is total and has a closed graph.
\end{claim}
\begin{proof}[Proof of \Cref{clm:counterexample-existence}]
Both simplices are compact. Continuity of the criteria implies that the maximizing correspondences have closed graphs, so \citet[Theorem~2]{kakutani1941} implies totality. The maximizing inequalities also remain valid along converging problems and assessments, yielding closedness.
\end{proof}

Now let us look at the absence of each axiom in turn:
\begin{enumerate}[wide=0pt]
\item\label{it:omit-rr} \hyperref[ax:RR]{\textbf{Invariance to randomization (\Cref*{ax:RR}).}} If $|A|\ge2$, select the uniform action distribution and every belief. If $|A|=1$, select all beliefs supported on models that maximize expected log likelihood, independently of payoffs. The rule is total and has a closed graph. In problems with more than one action, no pure assessment is selected, so \Cref{ax:PS,ax:SP,ax:PC,ax:HS} hold. Selection is independent of payoffs, so it also respects arbitrary convex combinations of payoff functions and every positive rescaling. \Cref{ax:IF,ax:SM,ax:EF} hold. To see the failure of \Cref{ax:RR}, take one model and two identical zero-payoff actions. Duplicate the first action. Choosing uniformly among the three actions induces probabilities $(2/3,1/3)$ on the original two actions, but the original problem selects only $(1/2,1/2)$, i.e., duplicating an action changes selection.
\item\label{it:omit-if} \hyperref[ax:IF]{\textbf{Invariance to irrelevant feedback (\Cref*{ax:IF}).}} Use the condition $\supp\alpha\subseteq\argmax_a\sum_\theta\nu_\theta m_{a\theta}$ in \eqref{eq:BN} and select beliefs supported on models maximizing $\sum_a\alpha_a H_{a\theta}$, where $H_{a\theta}\coloneqq\sum_z\sqrt{P_a(z)Q_{\theta a}(z)}$ is the Hellinger affinity. This rule is total and has a closed graph by \Cref{clm:counterexample-existence}.

For an action $c$ defined by the lottery $\lambda_c$ in \Cref{ax:RR}, both the true and model probabilities of $(a,z)$ are multiplied by $\lambda_c(a)$. The Hellinger affinity between the true distribution and model $\theta$'s distribution at $c$ is thus $\sum_a\lambda_c(a)H_{a\theta}$. The specified condition for choice implies \Cref{ax:PS,ax:PC,ax:SP,ax:HS}, as well as consistency under arbitrary convex combinations of payoff functions and invariance under positive rescaling.

With one action, the selected beliefs are those supported on maximizers of $H_{a\theta}$. Removing other models leaves these values unchanged. Moreover, $H_{a\theta}\le1$, with equality exactly when $Q_{\theta a}=P_a$, by the Cauchy-Schwarz inequality. Thus, \Cref{ax:SM,ax:EF} hold. In a problem with one action and zero payoffs, these belief sets also satisfy the properties concerning conditioning and model duplication in \Cref{cor:derived-coherences}.

To verify that \Cref{ax:IF} fails, take one zero-payoff action \(a\), $P=(1/2,1/2,0)$, $Q_1=(0.4,0.1,0.5)$, and $Q_2=(0.2,0.3,0.5)$. The values $H_{a1}$ and $H_{a2}$ are respectively $0.670820\ldots$ and $0.703526\ldots$. Append a binary report whose true outcome is always ``pass.'' Both models predict ``pass'' with conditional probabilities $0.81,0.04,0.5$ at the three original outcomes. After adding the report, these values become $0.447213\ldots$ and $0.362064\ldots$, so the rule selects model 1 after selecting model 2 in the original problem. All forecasts remain strictly positive and the premise of \Cref{ax:IF} holds.
\item\label{it:omit-ps} \hyperref[ax:PS]{\textbf{Payoff symmetry (\Cref*{ax:PS}).}} Set $\psi(\nu)_\theta\coloneqq\nu_\theta^2/\sum_\phi\nu_\phi^2$. Replace $\nu$ by $\psi(\nu)$ in the choice condition in \eqref{eq:BN}, and retain its fitting condition. By \Cref{clm:counterexample-existence}, the rule is total and has a closed graph. It satisfies \Cref{ax:RR,ax:IF,ax:PC,ax:SP,ax:HS}. Selection also respects arbitrary convex combinations of payoff functions and positive rescaling. In a problem with one action and zero payoffs, the selected beliefs are exactly those supported on models maximizing expected log likelihood. Hence \Cref{ax:SM,ax:EF} hold.

To verify that \Cref{ax:PS} fails, set $\nu=(1/4,3/4)$, $Q_{1a}=(0.8,0.2)$, $Q_{2a}=(0.4,0.6)$, and $P_a=(p,1-p)$ where $p=\log3/\log6$. The two models tie in expected log likelihood at $a$. The predictive distribution under $\nu$ is $(1/2,1/2)$, and the predictive distribution under $\psi(\nu)$ is $(0.44,0.56)$. Let $u_a=(0,1)$ and let another action $b$ pay $1/2$ for sure (its true distribution and both forecasts are $(1/2,1/2)$). The pure assessment at $a$ is selected because $0.56>1/2$. Exchanging the payoffs of the two outcomes changes expected payoff under $\psi(\nu)$ to $0.44<1/2$, so the rule no longer selects the assessment. This violates \Cref{ax:PS}.

\item\label{it:omit-pc} \hyperref[ax:PC]{\textbf{Midpoint payoff consistency (\Cref*{ax:PC}).}} Keep the fitting condition in \eqref{eq:BN}, but evaluate action $a$ by $\sum_{\theta,z}\nu_\theta Q_{\theta a}(z)u_a(z)^3$. By \Cref{clm:counterexample-existence}, the rule is total and has a closed graph. The cube is increasing, so \Cref{ax:SP} holds. Multiplying all payoffs by $t>0$ multiplies every action evaluation by $t^3$, leaving the maximizing actions unchanged; hence \Cref{ax:HS} holds, as does invariance under every positive rescaling. The rule also satisfies \Cref{ax:IF,ax:RR,ax:PS,ax:SM,ax:EF}. With one correctly specified model assigning probabilities $(1/2,1/2)$ at both actions, let $a$ pay $(0,2)$ and $b$ pay $3/2$ with certainty. The expected cubed utility of $a$ is four, exceeding $27/8$. The same assessment is selected after exchanging the two payoffs of $a$. Under the average of these two payoff functions, $a$ pays one with certainty and $b$ pays $3/2$, so the assessment at $a$ is no longer selected. \Cref{ax:PC} fails.

\item\label{it:omit-sp} \hyperref[ax:SP]{\textbf{Separated payoffs (\Cref*{ax:SP}).}} Keep the fitting condition in \eqref{eq:BN}, but \textit{minimize} subjective expected payoff. By \Cref{clm:counterexample-existence}, the rule is total and has a closed graph. It satisfies \Cref{ax:RR,ax:IF,ax:PS,ax:PC,ax:HS,ax:SM,ax:EF} with the action inequalities reversed. Selection is also consistent under arbitrary convex combinations of payoff functions and invariant under positive rescaling. In a correctly specified deterministic two-action problem paying zero and one, only the action paying zero is selected. \Cref{ax:SP} fails.

\item\label{it:omit-hs} \hyperref[ax:HS]{\textbf{Halving payoffs (\Cref*{ax:HS}).}} Let $V(\alpha,\nu;u,Q)\coloneqq\sum_{a,\theta}\alpha_a\nu_\theta m_{a\theta}$ be the expected payoff under policy $\alpha$ and belief $\nu$, where $m_{a\theta}=\sum_zQ_{\theta a}(z)u_a(z)$ as in \eqref{eq:payoffs-and-likelihoods}. Define
\begin{equation}\label{eq:regret} D(\alpha,\nu;P,Q)\coloneqq \max_\theta\sum_a\alpha_a\ell_{a\theta} -\sum_\theta\nu_\theta\sum_a\alpha_a\ell_{a\theta}\ge0. \end{equation}
The fitting condition ($\supp\nu\subseteq\argmax_\theta\sum_a\alpha_a\ell_{a\theta}$) in \eqref{eq:BN} is equivalent to $D=0$. Define $F^{\rm stakes}(u,P,Q)$ to consist of all assessments $(\alpha,\nu)$ satisfying  the choice condition ($\supp\alpha\subseteq\argmax_a\sum_\theta\nu_\theta m_{a\theta}$) in \eqref{eq:BN} and
\begin{equation}\label{eq:stakes-rule} D=0\quad\text{or}\quad V\ge D. \end{equation}
\begin{lemma}\label{lem:stakes}
$F^{\rm stakes}$ is total, has a closed graph, and satisfies \Cref{ax:RR,ax:IF,ax:PS,ax:PC,ax:SP,ax:SM,ax:EF}. It differs from $\BN$ and violates \Cref{ax:HS} in a one-action problem.
\end{lemma}
In \Cref{lem:stakes}, a belief excluded when every payoff is zero becomes selected when utility is sufficiently high. In the one-action example I use, \Cref{ax:HS} would preserve selection of the belief at constant payoff $2^{-n}$ for every $n\ge0$. Since $2^{-n}\to0$, a rule with a closed graph would also select the belief when every payoff is zero, as in the proof of \Cref{clm:diagnostic-assessment}.

\item\label{it:omit-sm} \hyperref[ax:SM]{\textbf{Model-menu consistency (\Cref*{ax:SM}).}} For each policy $\alpha$, form the skew-symmetric matrix on models
\begin{equation}\label{eq:M} M_{\theta\phi}(\alpha)\coloneqq\sum_{a,z}\alpha_aP_a(z) \frac{Q_{\theta a}(z)^2-Q_{\phi a}(z)^2} {Q_{\theta a}(z)^2+Q_{\phi a}(z)^2}. \end{equation}
Let $W(M)$ consist of the distributions $\nu$ over models that maximize $\min_{\phi\in\Theta}\sum_\theta\nu_\theta M_{\theta\phi}$. These are the maximizing player's optimal mixed strategies in the zero-sum game with matrix $M$. Since $M$ is skew-symmetric, the minimax equality implies that its value is zero \citep[Theorem~3]{kakutani1941}. Select precisely the pairs satisfying the choice condition in \eqref{eq:BN} and $\nu\in W(M(\alpha))$. The row player's minimum payoff is continuous and concave in its mixture, so $W(M(\alpha))$ is nonempty and convex. By \Cref{clm:counterexample-existence}, the rule is total and has a closed graph. The matrix in \eqref{eq:M} depends only on likelihood ratios at each outcome, so \Cref{ax:IF} holds. For an action defined by a lottery, the probabilities cancel from each ratio, and the resulting matrix is the corresponding weighted average of the original matrices. This proves \Cref{ax:RR}. The choice condition implies \Cref{ax:PS,ax:PC,ax:SP,ax:HS}, consistency under arbitrary convex combinations of payoff functions, and invariance under positive rescaling. The probability assigned to a model can also be divided arbitrarily among identical copies without changing selection.

For \Cref{ax:EF}, suppose $Q_{\theta^*}=P$. Since every model distribution has full support, $P\in\Delta^\circ(Z)$. Put $t_z\coloneqq Q_\phi(z)/P(z)$. Then
\begin{equation}\label{eq:true-matrix} M_{\theta^*\phi} =\sum_zP(z)\frac{1-t_z^2}{1+t_z^2} =\sum_zP(z)\frac{t_z(t_z-1)^2}{1+t_z^2}>0 \end{equation}
for an incorrect forecast, where the second equality uses $\sum_zP(z)(1-t_z)=0$. Any maximizing row strategy excludes incorrect models when compared with the column corresponding to the correct model, so \Cref{ax:EF} holds.

Consider a problem with one action, zero payoffs, and
\begin{equation}\label{eq:menu-cycle-forecasts} P \coloneqq\left(\frac{1}{3},\frac{1}{3},\frac{1}{3}\right), \quad Q_1 \coloneqq\left(\frac{4}{7},\frac{2}{7},\frac{1}{7}\right), \quad  Q_2 \coloneqq\left(\frac{2}{7},\frac{1}{7},\frac{4}{7}\right), \quad \text{\&} \quad Q_3 \coloneqq\left(\frac{1}{7},\frac{4}{7},\frac{2}{7}\right). \end{equation}
Substituting \eqref{eq:menu-cycle-forecasts} into \eqref{eq:M} yields $M_{12}=M_{23}=M_{31}=\frac13(\frac35+\frac35-\frac{15}{17})=9/85$. With all three models available, the unique maximizing belief is $(1/3,1/3,1/3)$. After model 3 is removed, the unique maximizing belief assigns probability one to model 1. Thus $C_F(P,Q)=\{1,2,3\}$, whereas $C_F(P,Q|_{\{1,2\}})=\{1\}$. This violates \Cref{ax:SM}.
\item\label{it:omit-ef} \hyperref[ax:EF]{\textbf{Binary exact fit (\Cref*{ax:EF}).}} Select every pair satisfying the choice condition in \eqref{eq:BN}, with no restriction on model fit. Choose any belief and an optimal action to establish totality. The choice inequalities define a closed graph. The transformations in \Cref{ax:RR,ax:IF,ax:PS,ax:PC,ax:HS} preserve selection under those inequalities, and expected-payoff maximization implies \Cref{ax:SP}. In a problem with one action and zero payoffs, every belief is selected. Thus \Cref{ax:SM} holds, as do the properties concerning conditioning and model duplication in \Cref{cor:derived-coherences}. An incorrect model may receive positive probability when a correct one is available, violating \Cref{ax:EF}.
\item\label{it:omit-totality} \textbf{Totality.} The rule $F(E)=\emptyset$ for every $E$ has a closed graph and satisfies all eight axioms. The sets $C_F(P,Q)$ are empty, so the premise of \Cref{ax:SM} is always false and \Cref{ax:EF} is vacuous.
\item \label{it:omit-closedness} \textbf{Closed graph.} Select precisely the Berk-Nash assessments with
\begin{equation}\label{eq:fullfit} \supp\nu=\argmax_\theta\sum_a\alpha_a\ell_{a\theta}. \end{equation}
This rule selects only those Berk-Nash assessments that assign positive probability to every model maximizing expected log likelihood. By \citet[Theorem~1]{norde1999}, the finite bimatrix game with payoff matrices $(m,\ell)$ has a quasi-strict equilibrium. Its row strategies are actions, its column strategies are models, and the two players' payoffs are $m$ and $\ell$. At such an equilibrium every pure best response receives positive probability, so each player's support equals its set of pure best responses. The equilibrium is a Berk-Nash assessment and satisfies \eqref{eq:fullfit}; hence, the rule is total. The transformations in \Cref{ax:RR,ax:IF} preserve the set of models maximizing expected log likelihood at a given policy. The choice condition in \eqref{eq:BN} implies \Cref{ax:PS,ax:PC,ax:SP,ax:HS}, consistency under arbitrary convex combinations of payoff functions, and invariance under positive rescaling.

In a problem with one action and zero payoffs, every selected belief assigns positive probability to exactly the models maximizing expected log likelihood. Thus $C_F(P,Q)$ is the set of all maximizing models, and the rule satisfies \Cref{ax:SM,ax:EF}. These belief sets also satisfy \eqref{eq:belief-menu}. With one action and two identical models, beliefs $(1-1/n,1/n)$ are selected for $n\ge2$, but their limit $(1,0)$ is not. Thus the rule does not have a closed graph. It does not permit arbitrary division of a model's probability among identical copies, because every maximizing copy must receive positive probability.
\end{enumerate}

\begin{proof}[Proof of \Cref{prop:independence}]
Each counterexample for an axiom is total, has a closed graph, and satisfies the other seven axioms. All counterexamples except the one for \Cref{ax:PC} also satisfy consistency under arbitrary convex combinations of payoff functions, and all except the one for \Cref{ax:HS} are invariant under positive rescaling. \Cref{it:omit-totality,it:omit-closedness} establish the independence of totality and the closed-graph property.
\end{proof}

\section{Discussion}\label{sec:interpretation}

\citet[Theorem~1]{brandl2024} characterizes Nash equilibrium among total solution concepts satisfying consequentialism, consistency, and rationality, which find natural mirrors among my axioms. Consequentialism requires that duplicating an action leave selection unchanged. \Cref{ax:RR} includes this requirement for copies that preserve the true and model distributions, and extends it to lotteries over actions. Consistency preserves a strategy profile selected in two games under every weighted average of their payoffs. Mine, \Cref{ax:PC}, requires only the equal-weight average, at a fixed belief and when an action is chosen with certainty. Rationality requires positive probability on an action that is strictly better than every other action, whatever the other players choose. Here, \Cref{ax:SP} rules out choosing an action with certainty if its best payoff is below another action's worst payoff.

In correlated equilibrium, a recommendation can convey information about what the other players will do. \citet[Definition~2]{brandl2026} distinguishes duplicating actions while preserving this information from duplications that change it. My observation requirement in \Cref{ax:RR} serves a related purpose: the decision maker observes the action drawn and can therefore compare each outcome with the forecasts for that action.

Here each model comes with a forecast $Q_\theta$, and the question is which models the rule should retain given the true distribution $P$. \citet{gilboa2003} instead begin with a reasoner's rankings of theories after different observations and derive forecasts under which theories with higher likelihoods rank higher. As \citet[Theorem~1]{gilboa2010} and the discussion that follows explain, these forecasts need not agree with forecasts specified in advance. \Cref{prop:inference} characterizes the rule that retains exactly the models maximizing expected log likelihood under $P$.

\citet[Theorem~3]{wang2024} also derive forecasts from choices among theories. Their rule can retain a theory even when another assigns a higher likelihood to the data: its likelihood need only be at least a fixed fraction of the supremum of likelihoods among the theories being considered. This can violate model-menu consistency, the first property in \secref{sec:model-selection}: after all but two theories are removed, a previously rejected theory may become selected even though the other theory was already selected.

My characterization concerns a single decision maker and does not specify a learning process. For the relation between Berk-Nash equilibrium and limits of Bayesian learning, see \citet[\S4]{esponda2016} and \citet[\S8]{esponda_pouzo_yamamoto2021}.

\bibliography{hpm}

\appendix

\section{Proof of the model-selection theorem \texorpdfstring{(\Cref{prop:inference})}{thm}}\label{sec:inference-proof}

\begin{lemma}\label{lem:common-refinement}
Let $X$ and $X'$ be finite nonempty sets of outcomes, and let $P\in\Delta(X)$ and $P'\in\Delta(X')$ be true distributions. Suppose that, for some $S\subseteq X\cap X'$, both distributions assign probability $p_s>0$ to each $s\in S$ and probability zero outside $S$. Let $Q=(Q_\theta)_{\theta\in\Theta}$ and $Q'=(Q'_\theta)_{\theta\in\Theta}$, where $Q_\theta\in\Delta^\circ(X)$ and $Q'_\theta\in\Delta^\circ(X')$ are the distributions specified by model $\theta$. If
\begin{equation}\label{eq:proportional-columns} Q'_\theta(s)=c_sQ_\theta(s)\quad(s\in S,\ \theta\in\Theta) \end{equation}
for positive numbers $c_s$ (\(s \in S\)), then $C(P,Q)=C(P',Q')$ for every rule satisfying \eqref{eq:feedback-invariance}.
\end{lemma}

\begin{proof}
Choose $b_s\in(0,\min\{1,c_s\})$ and set $\beta_{\theta s}\coloneqq b_sQ_\theta(s)$ and $\beta_{\theta x}\coloneqq0$ for outcomes outside $S$. Denote
\[r_\theta(x)\coloneqq Q_\theta(x)-\beta_{\theta x},\quad r'_\theta(x')\coloneqq Q'_\theta(x')-\beta_{\theta x'} \quad \text{\&} \quad D_\theta\coloneqq1-\sum_{s\in S}\beta_{\theta s}>0.\]
Let $W$ consist of one observation $g_s$ for each $s\in S$ and one observation $b_{xx'}$ for each pair $(x,x')\in X\times X'$. Define the true distribution $\widetilde P$ on $W$ to assign probability $p_s$ to $g_s$ and zero to every $b_{xx'}$. For each model $\theta$, define
\[\widetilde Q_\theta(g_s)\coloneqq\beta_{\theta s} \qquad \text{\&} \qquad \widetilde Q_\theta(b_{xx'})\coloneqq\frac{r_\theta(x)r'_\theta(x')}{D_\theta}.\]

Write $\widetilde Q=(\widetilde Q_\theta)_{\theta\in\Theta}$. Define functions $\rho\colon W\to X$ and $\rho'\colon W\to X'$ by
\begin{equation}\label{eq:common-refinement-maps} \rho(g_s)=\rho'(g_s)=s,\qquad \rho(b_{xx'})=x, \qquad \text{\&} \qquad \rho'(b_{xx'})=x'. \end{equation}
They are onto. Since $\sum_{x'\in X'}r'_\theta(x')=D_\theta$,
\begin{equation}\label{eq:common-refinement-model-marginals} \sum_{w\in\rho^{-1}(x)}\widetilde Q_\theta(w) =\beta_{\theta x} +\frac{r_\theta(x)}{D_\theta} \sum_{x'\in X'}r'_\theta(x') =Q_\theta(x). \end{equation}
Similarly, $\sum_{w\in(\rho')^{-1}(x')}\widetilde Q_\theta(w)=Q'_\theta(x')$. The corresponding sums for $\widetilde P$ equal $P(x)$ and $P'(x')$, since both \(\rho\) and \(\rho'\) send $g_s$ to $s$ and $\widetilde P$ assigns probability $p_s$ to $g_s$ and zero elsewhere.

For each $s\in S$, model $\theta$'s conditional probabilities of $g_s$ given $s$, under $\rho$ and $\rho'$ respectively, are
\begin{equation}\label{eq:common-refinement-conditional-agreement} \frac{\widetilde Q_\theta(g_s)}{Q_\theta(s)}=b_s \qquad \text{\&} \qquad \frac{\widetilde Q_\theta(g_s)}{Q'_\theta(s)}=\frac{b_s}{c_s}, \end{equation}
where the second equality uses \eqref{eq:proportional-columns}. These probabilities are independent of $\theta$, and the $g_s$ are the only observations with positive true probability. Applying \eqref{eq:feedback-invariance} to $\rho$ and $\rho'$ thus yields $C(P,Q)=C(\widetilde P,\widetilde Q)=C(P',Q')$.
\end{proof}

\begin{lemma}\label{lem:calibration-chains}
Fix $p\in\Delta^\circ(S)$ on a finite set $S$ with $|S|\ge2$. For every vector $d\in\mathbb R^S$ with $\sum_sp_sd_s>0$, there exist a positive integer $n$ and distributions $q,r\in\Delta^\circ(S)$, both different from $p$, such that
\begin{equation}\label{eq:decompose} d_s=n\left(\log\frac{p_s}{q_s}+\log\frac{p_s}{r_s}\right) \quad(s\in S). \end{equation}
\end{lemma}
\begin{proof}
\begin{claim}\label{clm:small-displacement}
There exists $n$ with $\sum_sp_s \mathrm{e}^{-d_s/n}<1$.
\end{claim}
\begin{proof}
Define $f(t)\coloneqq\sum_sp_s\mathrm{e}^{-td_s}$. Directly, $f(0)=1$ and $f'(0)=-\sum_sp_sd_s<0$, so $f(t)<1$ for all sufficiently small $t>0$. Thus, $\sum_sp_s\mathrm{e}^{-d_s/n}=f(1/n)<1$ for all sufficiently large positive integers $n$.
\end{proof}

Fix $n$ as in \Cref{clm:small-displacement}, and set $c_s\coloneqq \mathrm{e}^{-d_s/n}$ and $L\coloneqq\sum_sp_s\sqrt{c_s}$. The Cauchy-Schwarz inequality implies $L^2\le\sum_sp_sc_s<1$.

\begin{claim}\label{clm:calibration-vectors}
There exist distributions $q,r\in\Delta^\circ(S)$, both different from $p$, that satisfy \eqref{eq:decompose}.
\end{claim}
\begin{proof}[Proof of \Cref{clm:calibration-vectors}]
Fix $j\in S$. For $0\le t<1$, define $q(t)$ by
\[q_s(t)\coloneqq(1-t)\frac{p_s\sqrt{c_s}}{L} +t\mathbf 1\{s=j\} \quad(s\in S),\]
and set $F(t)\coloneqq\sum_s p_s^2c_s/q_s(t)$. The coordinates of $q(t)$ are positive and sum to one. Moreover, $F(0)=L^2<1$. As $t\uparrow1$, $q_s(t)\to0$ for every $s\ne j$. Since $|S|\ge2$, $F(t)\to\infty$. The function $F$ is continuous on $[0,1)$, so the intermediate value theorem implies that $F(t^*)=1$ for some $t^*\in(0,1)$.

Set $q=q(t^*)$ and $r_s\coloneqq p_s^2c_s/q_s$. Each $r_s$ is positive, and $\sum_s r_s=F(t^*)=1$, so $r\in\Delta^\circ(S)$. For every $s\in S$,
\[\log\frac{p_s}{q_s}+\log\frac{p_s}{r_s} =-\log c_s =\frac{d_s}{n},\]
which proves \eqref{eq:decompose}. If $q=p$, then $r_s=p_sc_s$; and if $r=p$, then $q_s=p_sc_s$. Either case contradicts $\sum_sp_sc_s<1$, since both $q$ and $r$ sum to one.
\end{proof}

The integer $n$ from \Cref{clm:small-displacement} and the distributions from \Cref{clm:calibration-vectors} satisfy the conclusion of \Cref{lem:calibration-chains}.
\end{proof}

\begin{lemma}\label{lem:strict-fit}
Every model-selection rule satisfying \eqref{eq:feedback-invariance}, \eqref{eq:model-menu}, and \eqref{eq:exact-fit} satisfies
\begin{equation}\label{eq:fit-contained} C(P,Q)\subseteq \argmax_{\theta\in\Theta}\sum_{z\in Z}P(z)\log Q_\theta(z). \end{equation}
\end{lemma}

\begin{proof}
Fix $p\in\Delta^\circ(S)$, where $|S|\ge2$. For a finite nonempty set of models $J$ and vectors $x_j\in\mathbb R^S$, $j\in J$, choose $c>0$ such that $c\sum_{s\in S}\mathrm{e}^{x_j(s)}<1$ for every $j\in J$. Let $\eta\notin S$. Define the true distribution $\bar p$ and the distributions specified by models $\bar Q_j$ on $S\cup\{\eta\}$ by
\begin{equation}\label{eq:log-vector-distributions} \bar p(s) \coloneqq p_s, \quad \bar p(\eta) \coloneqq0, \quad \bar Q_j(s) \coloneqq c\mathrm{e}^{x_j(s)}, \quad \text{\&} \quad \bar Q_j(\eta)\coloneqq1-c\sum_{t\in S}\mathrm{e}^{x_j(t)} \quad(s\in S, j\in J). \end{equation}
Write $\bar Q\coloneqq(\bar Q_j)_{j\in J}$. Replacing $c$ by another constant $c'>0$ multiplies every model's probability at each $s\in S$ by $c'/c$. Thus, \Cref{lem:common-refinement} implies that $C(\bar p,\bar Q)$ is unchanged, provided all model probabilities remain positive.

\begin{claim}\label{clm:translated-comparisons}
For $J=\{\theta,\phi\}$, suppose that $q\in\Delta^\circ(S)$, $q\ne p$, and \(x_\phi(s)=x_\theta(s)+\log\frac{p_s}{q_s}\) \((s\in S)\). Then the distributions in \eqref{eq:log-vector-distributions} satisfy $C(\bar p,\bar Q)=\{\phi\}$.
\end{claim}

\begin{proof} Consider the true distribution $p$ on $S$ and the family $H=(H_j)_{j\in\{\theta,\phi\}}$ defined by $H_\theta\coloneqq q$ and $H_\phi\coloneqq p$. Since $q\ne p$, \eqref{eq:exact-fit} implies $C(p,H)=\{\phi\}$. For each $s\in S$, set $a_s\coloneqq c\mathrm{e}^{x_\theta(s)}/q_s>0$. Then \(\bar Q_\theta(s)=a_sq_s\) and \(\bar Q_\phi(s)=a_sp_s\) and so by \Cref{lem:common-refinement}, $C(\bar p,\bar Q)=\{\phi\}$.
\end{proof}

\begin{claim}\label{clm:finite-chain-comparison}
For $J=\{\theta,\phi\}$, if \(\sum_{s\in S}p_s\left(x_\phi(s)-x_\theta(s)\right)>0\), then the distributions in \eqref{eq:log-vector-distributions} satisfy $C(\bar p,\bar Q)=\{\phi\}$.
\end{claim}

\begin{proof}[Proof of \Cref{clm:finite-chain-comparison}]
Set $x\coloneqq x_\theta$ and $y\coloneqq x_\phi$. By \Cref{lem:calibration-chains}, there exist a positive integer $n$ and distributions $q,r\in\Delta^\circ(S)$, both different from $p$, such that $y-x=n(v+w)$, where
\[v(s)\coloneqq\log\frac{p_s}{q_s}, \qquad \text{\&} \qquad w(s)\coloneqq\log\frac{p_s}{r_s} \qquad(s\in S).\]
Define vectors $u_0,\ldots,u_{2n}$ by
\begin{equation}\label{eq:strict-fit-sequence} u_{2k} \coloneqq x+k(v+w) \quad (0\le k\le n), \qquad \text{\&} \qquad u_{2k+1}\coloneqq x+k(v+w)+v \quad (0\le k<n). \end{equation}

By \Cref{lem:common-refinement}, reducing the constant $c$ in \eqref{eq:log-vector-distributions} does not change the selection from the original two models $\theta,\phi$, so we may  choose $c>0$ small enough that $c\sum_{s\in S}\mathrm e^{u_i(s)}<1$ for every $i=0,\ldots,2n$.

Denote the $2n-1$ additional models by $\theta_1,\ldots,\theta_{2n-1}$. Write $\theta_0\coloneqq\theta$, $\theta_{2n}\coloneqq\phi$, and $J\coloneqq\{\theta_0,\ldots,\theta_{2n}\}$. For each $i=0,\ldots,2n$, define the distribution of model $\theta_i$ on $S\cup\{\eta\}$ by
\begin{equation}\label{eq:strict-fit-chain-distributions} \bar Q^J_{\theta_i}(s) \coloneqq c\mathrm e^{u_i(s)} \quad (s\in S), \qquad \text{\&} \qquad \bar Q^J_{\theta_i}(\eta)\coloneqq1-c\sum_{s\in S}\mathrm e^{u_i(s)}. \end{equation}
Every $\bar Q^J_{\theta_i}$ has full support on $S\cup\{\eta\}$. Write $\bar Q^J\coloneqq(\bar Q^J_j)_{j\in J}$. The true distribution is $\bar p$ as specified in \eqref{eq:log-vector-distributions}.

Since $u_0=x=x_\theta$ and $u_{2n}=y=x_\phi$, \eqref{eq:strict-fit-chain-distributions} implies $\bar Q^J|_{\{\theta,\phi\}}=\bar Q$.

We first determine which model is selected when only $\theta_i$ and $\theta_{i+1}$ are available. For each $k=0,\ldots,n-1$, \eqref{eq:strict-fit-sequence} implies
\[u_{2k+1}(s)-u_{2k}(s)=\log\frac{p_s}{q_s}, \qquad \text{\&} \qquad u_{2k+2}(s)-u_{2k+1}(s)=\log\frac{p_s}{r_s} \quad(s\in S).\]
Applying \Cref{clm:translated-comparisons} to $\theta_i$ and $\theta_{i+1}$, with $q$ when $i$ is even and $r$ when $i$ is odd, we obtain
\begin{equation}\label{eq:strict-fit-adjacent-comparisons} C\left(\bar p,\bar Q^J|_{\{\theta_i,\theta_{i+1}\}}\right) =\{\theta_{i+1}\} \qquad(0\le i<2n). \end{equation}

Now suppose $\theta_i\in C(\bar p,\bar Q^J)$ for some $i<2n$. Then $C(\bar p,\bar Q^J)\cap\{\theta_i,\theta_{i+1}\}$ contains $\theta_i$ and is thus nonempty. The premise of \eqref{eq:model-menu} holds, so
\[C\left(\bar p,\bar Q^J|_{\{\theta_i,\theta_{i+1}\}}\right) = C(\bar p,\bar Q^J)\cap\{\theta_i,\theta_{i+1}\}.\]
The right-hand side contains $\theta_i$, whereas \eqref{eq:strict-fit-adjacent-comparisons} states that the left-hand side contains only $\theta_{i+1}$. This is a contradiction.

This argument excludes every model $\theta_0,\ldots,\theta_{2n-1}$ from $C(\bar p,\bar Q^J)$. Since the rule selects a nonempty set, the only remaining possibility is $C(\bar p,\bar Q^J)=\{\theta_{2n}\}=\{\phi\}$.

Finally, the original pair $\{\theta,\phi\}$ contains the selected model $\phi$. Applying \eqref{eq:model-menu} once more, now with $N=\{\theta,\phi\}$, yields \(C(\bar p,\bar Q)=C\left(\bar p,\bar Q^J|_{\{\theta,\phi\}}\right)=C(\bar p,\bar Q^J)\cap\{\theta,\phi\}=\{\phi\}\). \end{proof}

We now show that every model in $C(P,Q)$ maximizes expected log likelihood. Suppose first that $S\coloneqq\supp P$ has at least two elements. For $s\in S$, set $p_s\coloneqq P(s)$ and $x_\theta(s)\coloneqq\log Q_\theta(s)$. Define $\bar p$ and $\bar Q$ by \eqref{eq:log-vector-distributions} with $J=\Theta$. Since $\bar Q_\theta(s)=cQ_\theta(s)$ for every $s\in S$ and $\theta\in\Theta$, \Cref{lem:common-refinement} implies $C(P,Q)=C(\bar p,\bar Q)$.

If model $\theta$ has strictly lower expected log likelihood than model $\phi$, then $\sum_{s\in S}p_s(x_\phi(s)-x_\theta(s))>0$. Thus, \Cref{clm:finite-chain-comparison} implies\(C\left(\bar p,\bar Q|_{\{\theta,\phi\}}\right)=\{\phi\}\). By \eqref{eq:model-menu}, $\theta$ cannot belong to $C(\bar p,\bar Q)$, as its selection would imply that it is also selected from $\{\theta,\phi\}$. Accordingly, every model in $C(P,Q)$ maximizes expected log likelihood.

If $\supp P$ has one element, take $W\coloneqq Z\times\{0,1\}$ and define \(\widehat P(z,b)\coloneqq\frac12P(z)\), and \(\widehat Q_\theta(z,b)\coloneqq\frac12Q_\theta(z)\) \((z\in Z,\ b\in\{0,1\})\). Write $\widehat Q\coloneqq(\widehat Q_\theta)_{\theta\in\Theta}$. For the function $\rho(z,b)\coloneqq z$, every model assigns conditional probability $1/2$ to each of $(z,0)$ and $(z,1)$ given $z$. Therefore, \eqref{eq:feedback-invariance} implies $C(\widehat P,\widehat Q)=C(P,Q)$.

The support of $\widehat P$ has two elements, and
\[\sum_{(z,b)\in W}\widehat P(z,b)\log\widehat Q_\theta(z,b) = \sum_{z\in Z}P(z)\log Q_\theta(z)-\log2,\]
and so the result for true distributions with support of size at least two proves \eqref{eq:fit-contained} in this case as well.
\end{proof}

By \eqref{eq:inference-target}, renaming models or outcomes does not change which models the rule selects. The rule selects every copy of a model that maximizes expected log likelihood. It also satisfies \eqref{eq:model-menu} for every nonempty $N\subseteq\Theta$.

\begin{corollary}\label{cor:evidence-pooling}
Fix forecasts $Q$. If $C(P,Q)\cap C(P',Q)\ne\emptyset$, then for $0<\lambda<1$, \(C(\lambda P+(1-\lambda)P',Q)=C(P,Q)\cap C(P',Q)\).
\end{corollary}
\begin{proof}
Choose a model that maximizes expected log likelihood under both $P$ and $P'$. Its expected log likelihood minus that of any other model is nonnegative under each distribution. The weighted average of these two differences is zero exactly when both are zero, since both weights are positive. Apply \Cref{prop:inference}.
\end{proof}

By \Cref{lem:strict-fit}, every selected model maximizes expected log likelihood. To prove \Cref{prop:inference}, we next show that every maximizer is selected and then verify the converse. The first step uses the assumption that $C$ has a closed graph.

\begin{proof}[Proof of \Cref{prop:inference}]
First suppose that $C$ has a closed graph and satisfies \eqref{eq:feedback-invariance}, \eqref{eq:model-menu}, and \eqref{eq:exact-fit}.

\begin{claim}\label{clm:all-fitting-ties}
Every model that maximizes expected log likelihood belongs to $C(P,Q)$.
\end{claim}

\begin{proof}[Proof of \Cref{clm:all-fitting-ties}]
Set $L_\theta\coloneqq\sum_{z\in Z}P(z)\log Q_\theta(z)$ for each $\theta\in\Theta$, and fix $\theta^*\in\argmax_{\theta\in\Theta}L_\theta$. Let $W\coloneqq Z\times\{0,1\}$, and for $0\le\varepsilon<1/2$, define distributions on $W$ by
\begin{equation}\label{eq:tie-breaking-distributions} \begin{split} \widehat P(z,b) &\coloneqq\frac12P(z) \quad (b\in\{0,1\}) \quad  \text{\&} \quad \widehat Q_{\theta^*}^{\varepsilon}(z,b) \coloneqq\frac12Q_{\theta^*}(z) \quad (b\in\{0,1\}); \quad \text{\&}\\ \widehat Q_\theta^{\varepsilon}(z,0) &\coloneqq\left(\frac12+\varepsilon\right)Q_\theta(z) \quad (\theta\ne\theta^*) \quad \text{\&} \quad \widehat Q_\theta^{\varepsilon}(z,1) \coloneqq\left(\frac12-\varepsilon\right)Q_\theta(z) \quad (\theta\ne\theta^*). \end{split} \end{equation}
Write $\widehat Q^\varepsilon \coloneqq(\widehat Q_\theta^\varepsilon)_{\theta\in\Theta}$. Since each $Q_\theta$ has full support and $0\le\varepsilon<1/2$, \eqref{eq:tie-breaking-distributions} implies that each $\widehat Q_\theta^\varepsilon$ has full support on $W$. Moreover, the expected log likelihoods under $\widehat P$ are \(\widehat L^\varepsilon_{\theta^*} \coloneqq \sum_{(z,b)\in W}\widehat P(z,b) \log\widehat Q^\varepsilon_{\theta^*}(z,b) =L_{\theta^*}-\log2\) and
\begin{equation}\label{eq:tie-breaking-log-likelihoods} \widehat L^\varepsilon_\theta \coloneqq \sum_{(z,b)\in W}\widehat P(z,b) \log\widehat Q^\varepsilon_\theta(z,b) =L_\theta-\log2+\frac12\log(1-4\varepsilon^2) \qquad(\theta\ne\theta^*). \end{equation}
For $0<\varepsilon<1/2$, the last term in \eqref{eq:tie-breaking-log-likelihoods} is strictly negative, so $\theta^*$ is the unique maximizer. By \Cref{lem:strict-fit}, $C(\widehat P,\widehat Q^\varepsilon)\subseteq\{\theta^*\}$, and so, since model-selection rules select at least one model, \(C(\widehat P,\widehat Q^\varepsilon)=\{\theta^*\}\) for \(0<\varepsilon<1/2\).

Take a sequence $\varepsilon_k\in(0,1/2)$ converging to zero. Then $(\widehat P,\widehat Q^{\varepsilon_k}) \rightarrow (\widehat P,\widehat Q^0)$, and every distribution in $\widehat Q^0$ has full support. Since $C$ has a closed graph, $\theta^*\in C(\widehat P,\widehat Q^0)$. At $\varepsilon=0$, \eqref{eq:tie-breaking-distributions} implies that $\widehat Q_\theta^0(z,b)=Q_\theta(z)/2$ for every $\theta\in\Theta$ and $(z,b)\in W$. Thus, conditional on the first coordinate being $z$, model $\theta$ assigns probability $\frac{Q_\theta(z)/2}{Q_\theta(z)}=\frac12$ to each of $(z,0)$ and $(z,1)$. Applying \eqref{eq:feedback-invariance} with $\rho(z,b)\coloneqq z$ produces $C(\widehat P,\widehat Q^0)=C(P,Q)$ and so $\theta^*\in C(P,Q)$.
\end{proof}

\begin{claim}\label{clm:inference-converse}
The rule in \eqref{eq:inference-target} selects a nonempty set, has a closed graph, and satisfies \eqref{eq:model-menu}, \eqref{eq:exact-fit}, and \eqref{eq:feedback-invariance}.
\end{claim}

\begin{proof}[Proof of \Cref{clm:inference-converse}]
Recall that \(L_\theta(P,Q)\coloneqq \sum_{z\in Z}P(z)\log Q_\theta(z)\), which is continuous wherever all model distributions have full support. Since $\Theta$ is finite and nonempty, a maximizer exists.

Suppose $(P^k,Q^k)\to(P,Q)$, every limiting model distribution has full support, and model $\theta$ is selected for every $k$. Then for every $k$ and every $\phi\in\Theta$, \(L_\theta(P^k,Q^k)\ge L_\phi(P^k,Q^k)\) and taking the limit of both sides yields $L_\theta(P,Q)\ge L_\phi(P,Q)$ for every $\phi\in\Theta$, so $\theta$ is selected at the limit. Accordingly, the rule has a closed graph.

If $N\subseteq\Theta$ has two elements and contains a maximizer over $\Theta$, the maximum over $N$ equals the maximum over $\Theta$. Hence, the maximizers over $N$ are exactly the maximizers over $\Theta$ that belong to $N$. This proves \eqref{eq:model-menu}.

For \eqref{eq:exact-fit}, suppose $\Theta=\{\theta^*,\phi\}$, $Q_{\theta^*}=P$, and $Q_\phi\ne P$. As $\log t\le t-1$, with equality only at $t=1$:
\[L_\phi(P,Q)-L_{\theta^*}(P,Q) =\sum_{z\in Z}P(z)\log\frac{Q_\phi(z)}{P(z)} <\sum_{z\in Z}\left(Q_\phi(z)-P(z)\right)=0,\]
so the rule selects only $\theta^*$.

Finally, \eqref{eq:feedback} and $\sum_{w\colon\rho(w)=z}\widehat P(w)=P(z)$ imply
\begin{equation}\label{eq:inference-feedback-log-likelihood} \sum_{w\in W}\widehat P(w)\log\widehat Q_\theta(w) = \sum_{z\in Z}P(z)\log Q_\theta(z) + \sum_{\substack{w\in W\\\widehat P(w)>0}} \widehat P(w)\log R_\theta(w\mid\rho(w)). \end{equation}
Suppose $R_\theta(w\mid\rho(w))$ is the same for every model whenever $\widehat P(w)>0$. Then each term $\widehat P(w)\log R_\theta(w\mid\rho(w))$ in the last sum in \eqref{eq:inference-feedback-log-likelihood} is the same for every model, and so replacing $(P,Q)$ with $(\widehat P,\widehat Q)$ adds the same number to every model's expected log likelihood. Consequently, a model maximizes expected log likelihood under $(\widehat P,\widehat Q)$ if and only if it does so under $(P,Q)$. Since the rule in \eqref{eq:inference-target} selects exactly these models, $C(\widehat P,\widehat Q)=C(P,Q)$, which proves \eqref{eq:feedback-invariance}.
\end{proof}

Together, \Cref{lem:strict-fit} and \Cref{clm:all-fitting-ties} deliver \eqref{eq:inference-target}, and \Cref{clm:inference-converse} proves the converse.
\end{proof}

\section{Proofs of the joint characterization lemmata}

\subsection{Proof of \texorpdfstring{\Cref{lem:inference-containment}}{lemma}}
\begin{proof}[Proof of \Cref{lem:inference-containment}]
\begin{claim}\label{clm:diagnostic-supports}
In a problem with one action and zero payoffs, every model in $C_F(P,Q)$ maximizes expected log likelihood.
\end{claim}
\begin{proof}[Proof of \Cref{clm:diagnostic-supports}]
Since $F$ is total, the set $C_F(P,Q)$ in \eqref{eq:SC} is nonempty and so $C_F$ is a model-selection rule. \Cref{ax:IF} implies \eqref{eq:feedback-invariance} for $C_F$, while \Cref{ax:SM,ax:EF} imply \eqref{eq:model-menu} and \eqref{eq:exact-fit}. Apply \Cref{lem:strict-fit}.
\end{proof}

For a policy $\alpha$, let $Z^\alpha\coloneqq\{(c,z)\colon c\in\supp\alpha,\ z\in Z_c\}$ and define distributions on $Z^\alpha$ by $P^\alpha(c,z)\coloneqq\alpha_cP_c(z)$ and $Q^\alpha_\theta(c,z)\coloneqq\alpha_cQ_{\theta c}(z)$. Write $Q^\alpha\coloneqq(Q^\alpha_\theta)_{\theta\in\Theta}$. Every $Q^\alpha_\theta$ has full support on $Z^\alpha$.

\begin{claim}\label{clm:diagnostic-assessment}
If $(\alpha,\nu)\in F(u,P,Q)$, then $\nu\in S_F(P^\alpha,Q^\alpha)$.
\end{claim}
\begin{proof}[Proof of \Cref{clm:diagnostic-assessment}]
Fix $(\alpha,\nu)\in F(u,P,Q)$. Add a new action $a$ that draws an original action $c$ according to $\alpha$. The decision maker observes both $c$ and its outcome $z$, and receives payoff $u_c(z)$. This action has distributions $P^\alpha,Q^\alpha$, and \Cref{ax:RR} implies that $(\delta_a,\nu)$ is selected in the expanded problem. By repeated application of \Cref{ax:HS}, this assessment remains selected when all payoffs are multiplied by $2^{-n}$. Taking $n\to\infty$ and using the closed graph of $F$, it remains selected when all payoffs are zero.

We next make every other action identical to $a$, leaving the distributions at $a$ unchanged. For each original action $b$, add an observation $y\in Z^\alpha$ to its outcome $z_b$, with true joint probability $P_b(z_b)P^\alpha(y)$ and model joint probability $Q_{\theta b}(z_b)Q^\alpha_\theta(y)$. Discarding $y$ recovers the original distributions at $b$. Since every such $b$ has probability zero under $\delta_a$, \Cref{ax:IF} preserves selection of $(\delta_a,\nu)$.

Now retain only $y$ at each original action $b$. Its true distribution is $P^\alpha$, and model $\theta$ specifies $Q^\alpha_\theta$. The preceding joint observation can be obtained by adding $z_b$ to $y$, with true conditional distribution $P_b$ and model conditional distribution $Q_{\theta b}$. Because all payoffs are zero and each $b$ remains unused, \Cref{ax:IF} again preserves selection of $(\delta_a,\nu)$.

All actions now have distributions $P^\alpha,Q^\alpha$ and zero payoffs. By \Cref{ax:RR}, selection of $(\delta_a,\nu)$ among these identical actions implies selection of $\nu$ in the corresponding problem with one action. Thus, $\nu\in S_F(P^\alpha,Q^\alpha)$.
\end{proof}
Now fix $(\alpha,\nu)\in F(u,P,Q)$. For each model $\theta$,
\[\sum_{(c,z)\in Z^\alpha}P^\alpha(c,z)\log Q^\alpha_\theta(c,z) =\sum_c\alpha_c\ell_{c\theta} +\sum_{c\in\supp\alpha}\alpha_c\log\alpha_c.\]
The last sum is the same for every model. By \Cref{clm:diagnostic-assessment}, every model in $\supp\nu$ belongs to $C_F(P^\alpha,Q^\alpha)$. Applying \Cref{clm:diagnostic-supports} delivers the lemma.
\end{proof}

\subsection{Proof of \texorpdfstring{\Cref{lem:choice-containment}}{lemma}}
\begin{proof}[Proof of \Cref{lem:choice-containment}]
Suppose $(\alpha,\nu)\in F(E)$ and $\alpha$ assigns positive probability to an action that does not maximize expected utility under $\nu$. Add an action $a$ that draws an original action $c$ according to $\alpha$. The decision maker observes both $c$ and its outcome $z$ and receives payoff $u_c(z)$. By \Cref{ax:RR}, $(\delta_a,\nu)$ is selected. The expected utility of $a$ is the $\alpha$-weighted average of the original actions' expected utilities, so some original action $b$ has strictly greater expected utility. We next make every payoff at each of these two actions close to its expected utility while preserving selection of $(\delta_a,\nu)$.

For each $d\in\{a,b\}$, write $p_d\coloneqq\overline Q_d^\nu$ and $\mu_d\coloneqq\sum_zp_d(z)u_d(z)$. Fix $n=2^k$. At each action $d$, partition $[0,1)$ into consecutive half-open intervals $I_{dz}$ of lengths $p_d(z)$. Split these intervals at every point obtained by translating a partition endpoint by $j/n$, modulo one, for $j=0,\ldots,n-1$.\footnote{This ensures that translation by $1/n$ sends each resulting interval onto another resulting interval of the same length.} Use the intervals of strictly positive length as new observations. Conditional on outcome $z$, assign each interval $J\subseteq I_{dz}$ probability $|J|/p_d(z)$ under the true distribution and every model, where $|J|$ denotes its length. Its payoff remains $u_d(z)$. Since these conditional probabilities are the same under every model, \Cref{ax:IF} preserves selection. The probability of $J$ under $\nu$ is $p_d(z)|J|/p_d(z)=|J|$.

For each $j=0,\ldots,n-1$, replace the payoff at each new interval $J$ by the payoff at $J+j/n$, modulo one, at both $a$ and $b$. Translated intervals have the same length and hence the same probability under $\nu$, so \Cref{ax:PS} implies that $(\delta_a,\nu)$ remains selected after each of these $n$ permutations of the payoffs. We average these payoffs in pairs and then repeat the procedure with the resulting averages. At each stage, \Cref{ax:PC} preserves selection of $(\delta_a,\nu)$. Since $n=2^k$, after $k$ rounds we obtain the average of all $n$ permuted payoffs. Write $v$ for these average payoffs.

For a new observation $J$ at action $d\in\{a,b\}$, the average payoff $v_d(J)$ depends on how many of its $n$ translates lie in each original interval $I_{dz}$. To count these translates, choose a point $x$ in the interior of $J$. The points $x+j/n$, modulo one, are equally spaced, so each original interval $I_{dz}$ contains either $\lfloor np_d(z)\rfloor$ or $\lceil np_d(z)\rceil$ of them. Each translated interval lies in the same original interval $I_{dz}$ as its translated point and thus has payoff $u_d(z)$. Thus the weight assigned to $u_d(z)$ in the average $v_d(J)$ differs from $p_d(z)$ by at most $1/n$, implying
\[|v_d(J)-\mu_d| \leq \frac{1}{n}\sum_{z\in Z_d}|u_d(z)|.\]
This bound holds at every new observation. Since $\mu_b>\mu_a$, taking $n$ sufficiently large makes every payoff at $b$ exceed every payoff at $a$. Selection of $(\delta_a,\nu)$ then contradicts \Cref{ax:SP}.
\end{proof}

\subsection{Proof of \texorpdfstring{\Cref{lem:no-refinement}}{lemma}}
\begin{proof}[Proof of \Cref{lem:no-refinement}]
Fix $(\alpha,\nu)\in\BN(E)$. Retain every original action and add a new action $a$ that draws an original action $c$ according to $\alpha$. The decision maker observes both $c$ and its outcome $z$ and receives payoff $u_c(z)$. For every $\theta\in\Theta$, $m_{a\theta}=\sum_{c\in A}\alpha_cm_{c\theta}$ and $\ell_{a\theta}=\sum_{c\in A}\alpha_c\ell_{c\theta}+\sum_{c\in\supp\alpha}\alpha_c\log\alpha_c$. Since the last sum does not depend on $\theta$, these identities imply that $(\delta_a,\nu)$ is a Berk-Nash assessment of the expanded problem. By \Cref{ax:RR}, it suffices to prove that $F$ selects $(\delta_a,\nu)$ in this problem.

Let $S\coloneqq\supp\nu$, $K\coloneqq\max_{\theta\in\Theta}\ell_{a\theta}$, and $d_\theta\coloneqq K-\ell_{a\theta}$. Then $d_\theta\geq0$ for every $\theta\in\Theta$ and, since $(\delta_a,\nu)$ is a Berk-Nash assessment, $d_\theta=0$ for every $\theta\in S$.

\begin{claim}\label{clm:unused-action-reports}
We can add a binary report at each action $b\ne a$, without changing expected payoffs, so that the resulting expected log likelihoods $\ell^D$ satisfy \(\ell^D_{b\theta}=\ell_{a\theta}-(m_{b\theta}-m_{a\theta})+c_b\) for constants $c_b$. Adding these reports preserves selection of $(\delta_a,\nu)$ under $F$.
\end{claim}
\begin{proof}[Proof of \Cref{clm:unused-action-reports}]
At each action $b\ne a$, add an independent binary report that is always zero under the true distribution. Define $\ell^D_{b\theta}$ by \(\ell^D_{b\theta}=\ell_{a\theta}-(m_{b\theta}-m_{a\theta})+c_b\), and set model $\theta$'s probability of zero equal to $\mathrm{e}^{\ell^D_{b\theta}-\ell_{b\theta}}$. A sufficiently negative $c_b$ places this probability in $(0,1)$ for every model. The report then changes expected log likelihood from $\ell_{b\theta}$ to $\ell^D_{b\theta}$. Its realization has no effect on payoffs. Since $b$ is unused, \Cref{ax:IF} implies that $(\delta_a,\nu)$ is selected after adding the report if and only if it was selected before.
\end{proof}

Replace $a$ by actions $a_j$, $j\in S$, each with the same outcomes, payoffs, true distribution, and model distributions as $a$. Retain every other action $b$. Define
\begin{equation}\label{eq:copy-weights} R_{j\theta}\coloneqq \begin{cases} \mathbf1_{\{j=\theta\}}/\nu_\theta,&\theta\in S,\\ 2,&\theta\notin S. \end{cases} \end{equation}
Let $\tau$ be the policy that assigns probability $\nu_j$ to $a_j$ and zero to every other action.

Choose $C>0$ so that $C+r>r^2/4$ for every value $r=R_{j\theta}$. For $\varepsilon>0$, add an observation $x\in\{1,2,3\}$ at each $a_j$, independent of the original outcome $z$ under the true distribution and every model. Under the true distribution, each value of $x$ has probability $1/3$. The decision maker observes both $z$ and $x$ and receives payoff $u_a(z)+2\varepsilon$ if $x=1$ and $u_a(z)-\varepsilon$ if $x=2$ or $3$. For each pair $j,\theta$, set $r\coloneqq R_{j\theta}$ and choose model $\theta$'s probabilities $q_1,q_2,q_3$ of the three values of $x$ at action $a_j$ so that
\begin{equation}\label{eq:perturbation-probabilities} q_1=\frac{1+\varepsilon r}{3},\qquad q_2+q_3=1-q_1 \qquad \text{\&} \qquad q_2q_3=\frac{\mathrm{e}^{-3\varepsilon^2(C+r)}}{27q_1}. \end{equation}
For all sufficiently small positive $\varepsilon$, positive $q_2,q_3$ exist and converge to $1/3$. To see this, fix $q_1$ and note that positive $q_2,q_3$ with sum $1-q_1$ can have any product in $(0,(1-q_1)^2/4]$. Since
\[\lim_{\varepsilon\downarrow0} -\frac{1}{3\varepsilon^2} \log \left(\frac{27q_1(1-q_1)^2}{4}\right) =\frac{r^2}{4}<C+r,\]
we have $\mathrm{e}^{-3\varepsilon^2(C+r)}/(27q_1)<(1-q_1)^2/4$ for all sufficiently small $\varepsilon>0$. Moreover, $q_2+q_3\to2/3$ and $q_2q_3\to1/9$, so $q_2,q_3\to1/3$. Write $E^\varepsilon$ for the resulting problem.

\begin{claim}\label{clm:unique-perturbed-assessment}
For all sufficiently small positive $\varepsilon$, $(\tau,\nu)$ is the unique Berk-Nash assessment of $E^\varepsilon$.
\end{claim}
\begin{proof}[Proof of \Cref{clm:unique-perturbed-assessment}]
By \eqref{eq:perturbation-probabilities},
\[\begin{aligned} m^\varepsilon_{a_j\theta} &=m_{a\theta}+\varepsilon(3q_1-1) =m_{a\theta}+\varepsilon^2R_{j\theta}, \qquad \text{\&}\\ \ell^\varepsilon_{a_j\theta} &=\ell_{a\theta}+\frac13\log(q_1q_2q_3) =\ell_{a\theta}-\log3-\varepsilon^2(C+R_{j\theta}). \end{aligned}\]
Define
\begin{equation}\label{eq:saddle-matrix} H_{a_j\theta}\coloneqq d_\theta+\varepsilon^2R_{j\theta}, \qquad \text{\&} \qquad H_{b\theta}\coloneqq d_\theta+m_{b\theta}-m_{a\theta}. \end{equation}
As \(\ell^D_{b\theta} =\ell_{a\theta}-(m_{b\theta}-m_{a\theta})+c_b\), we obtain
\[m^\varepsilon_{c\theta}-H_{c\theta} =m_{a\theta}-d_\theta \text{ for every action $c$}, \quad \ell^\varepsilon_{a_j\theta}+H_{a_j\theta} =K-\log3-\varepsilon^2C \quad \text{\&} \quad \ell^\varepsilon_{b\theta}+H_{b\theta} =K+c_b.\]
For a policy $\beta$ and a belief $\mu$, write
\[(H\mu)_c\coloneqq\sum_\theta\mu_\theta H_{c\theta}, \qquad (\beta H)_\theta\coloneqq\sum_c\beta_cH_{c\theta}, \qquad \text{\&} \qquad \beta H\mu\coloneqq\sum_{c,\theta}\beta_c\mu_\theta H_{c\theta}.\]
For any belief $\mu$, maximizing expected payoff over actions is equivalent to maximizing $(H\mu)_c$, since $m^\varepsilon_{c\theta}-H_{c\theta}$ does not depend on $c$. For any policy $\beta$, maximizing expected log likelihood over models is equivalent to minimizing $(\beta H)_\theta$, since $\ell^\varepsilon_{c\theta}+H_{c\theta}$ does not depend on $\theta$. Hence the Berk-Nash assessments of $E^\varepsilon$ are exactly the equilibria of the zero-sum game with payoff matrix $H$, where the maximizer  chooses an action and the minimizer chooses a model. Consequently,$(\beta,\mu)\in\BN(E^\varepsilon)$ if and only if \(\supp\beta\subseteq\argmax_c(H\mu)_c\) and \(\supp\mu\subseteq\operatorname*{arg\,min}_\theta(\beta H)_\theta\).

By \eqref{eq:copy-weights} and the fact that $d_\theta=0$ on $S$,
\[(\tau H)_\theta = \begin{cases} \varepsilon^2,&\theta\in S,\\ d_\theta+2\varepsilon^2,&\theta\notin S, \end{cases} \qquad (H\nu)_{a_j} =\varepsilon^2, \qquad \text{\&} \qquad (H\nu)_b =\sum_\theta\nu_\theta(m_{b\theta}-m_{a\theta}) \leq0.\]
The last inequality holds because $a$ maximizes expected utility under $\nu$. Hence, $(\tau,\nu)\in\BN(E^\varepsilon)$ and $\tau H\nu=\varepsilon^2$.

Now let $(\beta,\mu)\in\BN(E^\varepsilon)$. Its two optimality conditions imply $\tau H\mu\leq\beta H\mu\leq\beta H\nu$. Together with the preceding calculations,
\[\varepsilon^2 \leq\tau H\mu \leq\beta H\mu \leq\beta H\nu \leq\varepsilon^2,\]
and so \(\tau H\mu = \beta H\mu = \beta H\nu\). Since $(\tau H)_\theta>\varepsilon^2$ outside $S$, the left-hand equality requires $\mu_\theta=0$ outside $S$. Since $(H\nu)_b<\varepsilon^2$ at every other action $b$, the right-hand equality requires $\beta_b=0$ at all such actions.

For each $j\in S$, optimality of $\beta$ and $\mu$ now implies
\[\varepsilon^2\frac{\mu_j}{\nu_j} \leq\beta H\mu =\varepsilon^2 \leq\varepsilon^2\frac{\beta_{a_j}}{\nu_j}.\]
Thus, $\mu_j\leq\nu_j\leq\beta_{a_j}$ for every $j\in S$. Since each of these three distributions sums to one on $S$, $\mu_j=\nu_j=\beta_{a_j}$ for every $j\in S$ and so $(\beta,\mu)=(\tau,\nu)$.
\end{proof}

\begin{claim}\label{clm:restore-assessment}
The original assessment $(\alpha,\nu)$ belongs to $F(E)$.
\end{claim}
\begin{proof}[Proof of \Cref{clm:restore-assessment}]
By \Cref{clm:unique-perturbed-assessment}, $(\tau,\nu)$ is the unique Berk-Nash assessment of $E^\varepsilon$. Since $F$ is total and selects only Berk-Nash assessments, it selects $(\tau,\nu)$. As $\varepsilon\downarrow0$, the problems converge on the same finite sets, and every limiting model distribution has full support. Since $F$ has a closed graph, $(\tau,\nu)$ is selected in the limiting problem.

At the limit, every model assigns probability $1/3$ to each of the three report realizations at every copy, and the report has no effect on payoffs. Removing it preserves selection by \Cref{ax:IF}. Replacing the copies of $a$ by $a$ preserves selection by \Cref{ax:RR}. Every action $b\ne a$ still has probability zero under $\delta_a$. Removing the observations added at these actions in \Cref{clm:unused-action-reports} thus preserves selection of $(\delta_a,\nu)$ by \Cref{ax:IF}. Finally, \Cref{ax:RR} preserves selection when we remove the added action $a$ and recover the original assessment $(\alpha,\nu)$.
\end{proof}

Since $(\alpha,\nu)$ was arbitrary, \Cref{clm:restore-assessment} implies $\BN(E)\subseteq F(E)$, and $\BN(E)\supseteq F(E)$ by assumption.
\end{proof}

\section{Proof of \texorpdfstring{\Cref{lem:stakes}}{prop}}

\begin{proof}[Proof of \Cref{lem:stakes}]
\begin{claim}\label{clm:stakes-diagnostics}
The rule $F^{\rm stakes}$ is total, has a closed graph, and satisfies \Cref{ax:SM,ax:EF}.
\end{claim}
\begin{proof}[Proof of \Cref{clm:stakes-diagnostics}]
Every Berk-Nash assessment satisfies \eqref{eq:stakes-rule}, so $F^{\rm stakes}$ is total. The choice condition in \eqref{eq:BN} defines a closed set of problems and assessments. Both $\{D=0\}$ and $\{V\ge D\}$ are closed because $V$ and $D$ are continuous, so their union is closed. Hence, $F^{\rm stakes}$ has a closed graph.

When $u_a(z)=0$ for every $a\in A$ and $z\in Z_a$, $V=0$ and \eqref{eq:stakes-rule} is equivalent to $D=0$. In a problem with one action and zero payoffs, the rule thus selects exactly the beliefs supported on models maximizing expected log likelihood. By \Cref{clm:inference-converse}, these belief sets satisfy \Cref{ax:SM,ax:EF}. They also satisfy the properties concerning conditioning and model duplication in \Cref{cor:derived-coherences}.
\end{proof}

\begin{claim}\label{clm:stakes-coherence}
The rule $F^{\rm stakes}$ satisfies \Cref{ax:RR,ax:IF,ax:PS,ax:PC,ax:SP}.
\end{claim}
\begin{proof}[Proof of \Cref{clm:stakes-coherence}]
The lotteries in \Cref{ax:RR} preserve $V$ and $D$. They leave expected payoffs unchanged, and the added terms in expected log likelihood are the same for every model and cancel from $D$. Adding observations as in \Cref{ax:IF} also preserves both quantities at the assessment being considered. When $\alpha=\delta_a$, so that action $a$ is chosen with certainty, the permutations in \Cref{ax:PS} preserve every action's expected payoff under $\nu$ and leave $D$ fixed. For fixed $(\alpha,\nu,P,Q)$, changing payoffs does not change $D$. If $D=0$, the rule selects the assessment under every payoff function satisfying the condition $\supp\alpha\subseteq\argmax_a\sum_\theta\nu_\theta m_{a\theta}$ in \eqref{eq:BN}. If $D>0$, selection also depends on the linear inequality $V\ge D$. Thus, if an assessment is selected under two payoff functions, it remains selected under every convex combination of them, so \Cref{ax:PC} holds. The condition $\supp\alpha\subseteq\argmax_a\sum_\theta\nu_\theta m_{a\theta}$ in \eqref{eq:BN} implies \Cref{ax:SP}.
\end{proof}

\begin{claim}\label{clm:stakes-failure}
The rule $F^{\rm stakes}$ differs from $\BN$ and violates \Cref{ax:HS}.
\end{claim}
\begin{proof}[Proof of \Cref{clm:stakes-failure}]
Consider a problem with one action and constant payoff one. Let $P=(1/2,1/2)$, $Q_1=(1/2,1/2)$, and $Q_2=(9/10,1/10)$. At $\nu=\delta_2$, $D=\log(5/3)$, strictly between $1/2$ and one. Thus, the belief assigning probability one to model 2 is selected at payoff one, but not at payoff $1/2$. The assessment is not Berk-Nash, and \Cref{ax:HS} fails. The rule also fails invariance under positive rescaling.
\end{proof}

Together, \Cref{clm:stakes-diagnostics,clm:stakes-coherence,clm:stakes-failure} prove the proposition.
\end{proof}
\end{document}